\documentclass[aps,prl,twocolumn,superscriptaddress]{revtex4}

\usepackage{graphicx,subfigure}
\usepackage{amsfonts}

\pdfoutput=1

\begin{document}

\title{Homological characterizations of spiral defect chaos in
Rayleigh-B\'{e}nard convection}

\author{Kapilanjan Krishan}
\email[E-mail: ]{kapil@cns.physics.gatech.edu}
\affiliation{School of Physics, Georgia Institute of Technology,
Atlanta, Georgia, 30332, USA}

\author{Marcio Gameiro}
\email[E-mail: ]{gameiro@math.gatech.edu}
\affiliation{School of Mathematics, Georgia Institute of Technology,
Atlanta, Georgia, 30332, USA}

\author{Konstantin Mischaikow}
\email[E-mail: ]{mischaik@math.gatech.edu}
\affiliation{School of Mathematics, Georgia Institute of Technology,
Atlanta, Georgia, 30332, USA}

\author{Michael F. Schatz}
\email[E-mail: ]{mike.schatz@physics.gatech.edu}
\affiliation{School of Physics, Georgia Institute of Technology,
Atlanta, Georgia, 30332, USA}

\date{April 15, 2005}

\begin{abstract}
We use a quantitative topological characterization of complex dynamics to
measure geometric structures. This approach is used to analyze the weakly
turbulent state of spiral defect chaos in experiments on Rayleigh-B\'{e}nard
convection. Different attractors of spiral defect chaos are distinguished by their
homology. The technique reveals pattern asymmetries that are not revealed using
statistical measures.  In addition we observe global stochastic ergodicity for
system parameter values where locally chaotic dynamics has been observed previously.
\end{abstract}

\maketitle

Characterization of geometric structures (patterns) in physical systems can give
insight into underlying dynamics.  For simple cases, pattern characterization is
easily done with a few numbers (e.g. the lattice constants describing problems
with crystalline symmetries). However encoding patterns becomes more difficult as
patterns become more complex. Statistical approaches can be used to describe complex
patterns in dynamically evolving systems \cite{Morris_physicaD,Ecke_science};
however, a general methodology for extracting geometric signatures from complex
patterns has been lacking. 

\begin{figure}[!hbt]%
\begin{center}%
{\includegraphics[width=4.2cm]{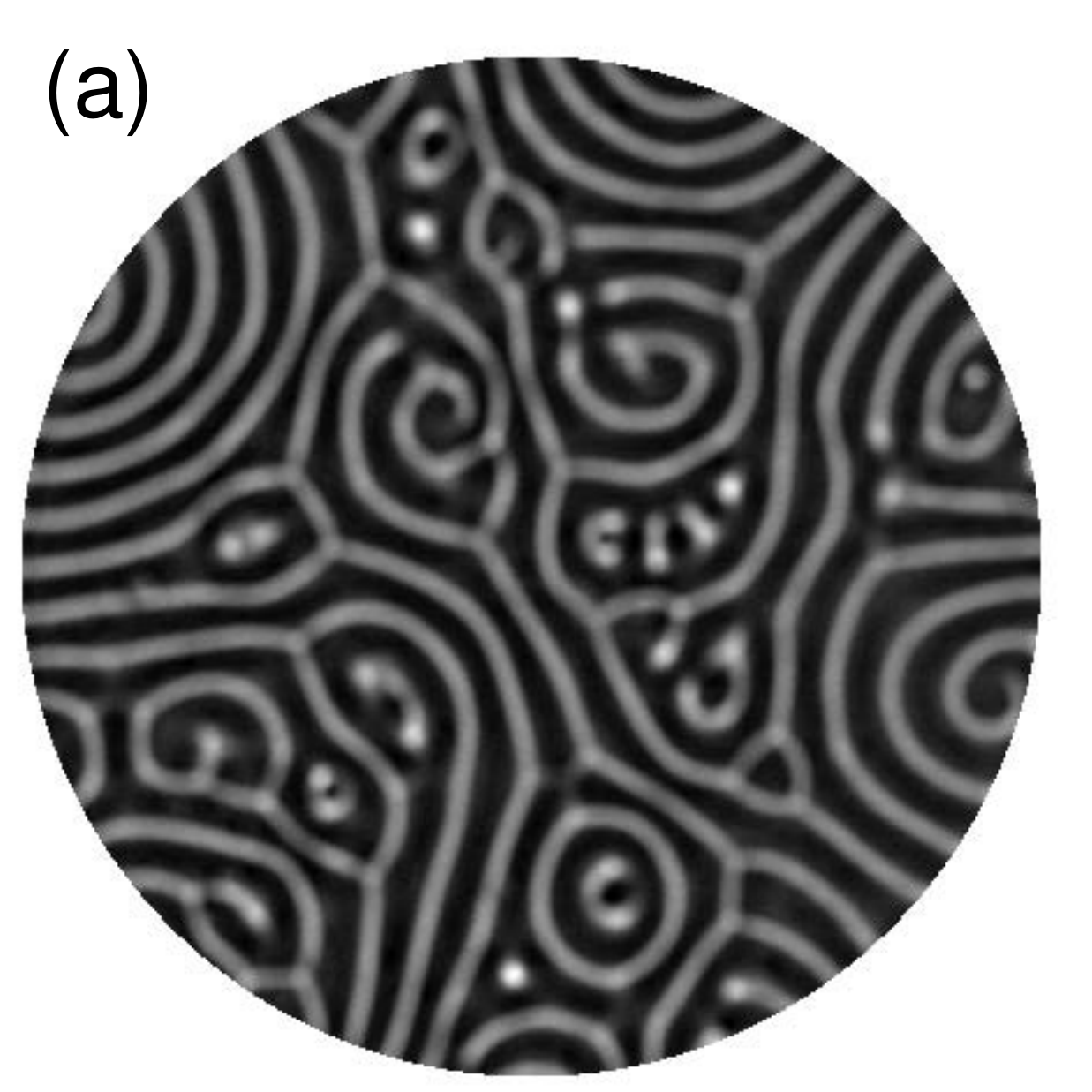}}
{\includegraphics[width=4.2cm]{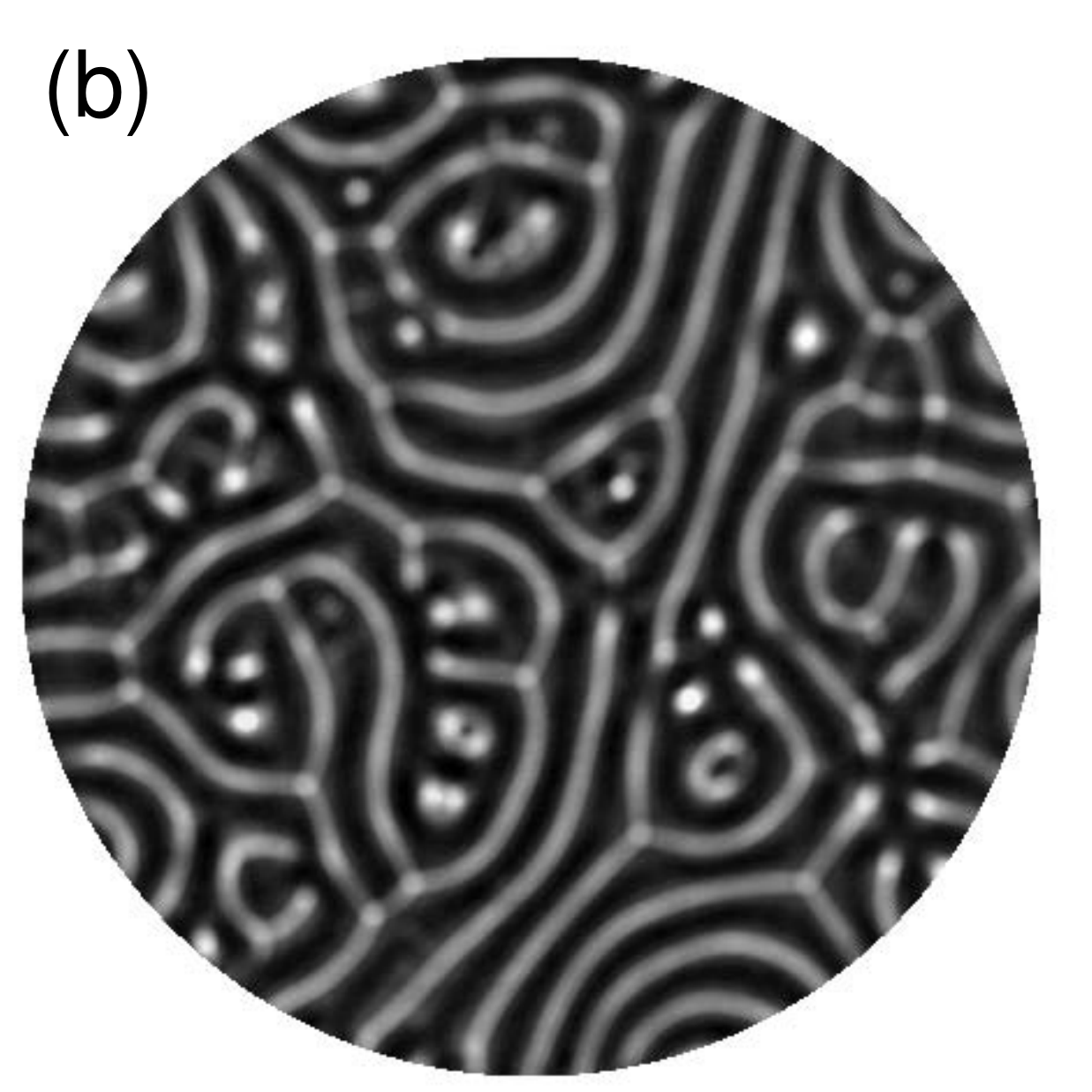}} \\
{\includegraphics[width=4.2cm]{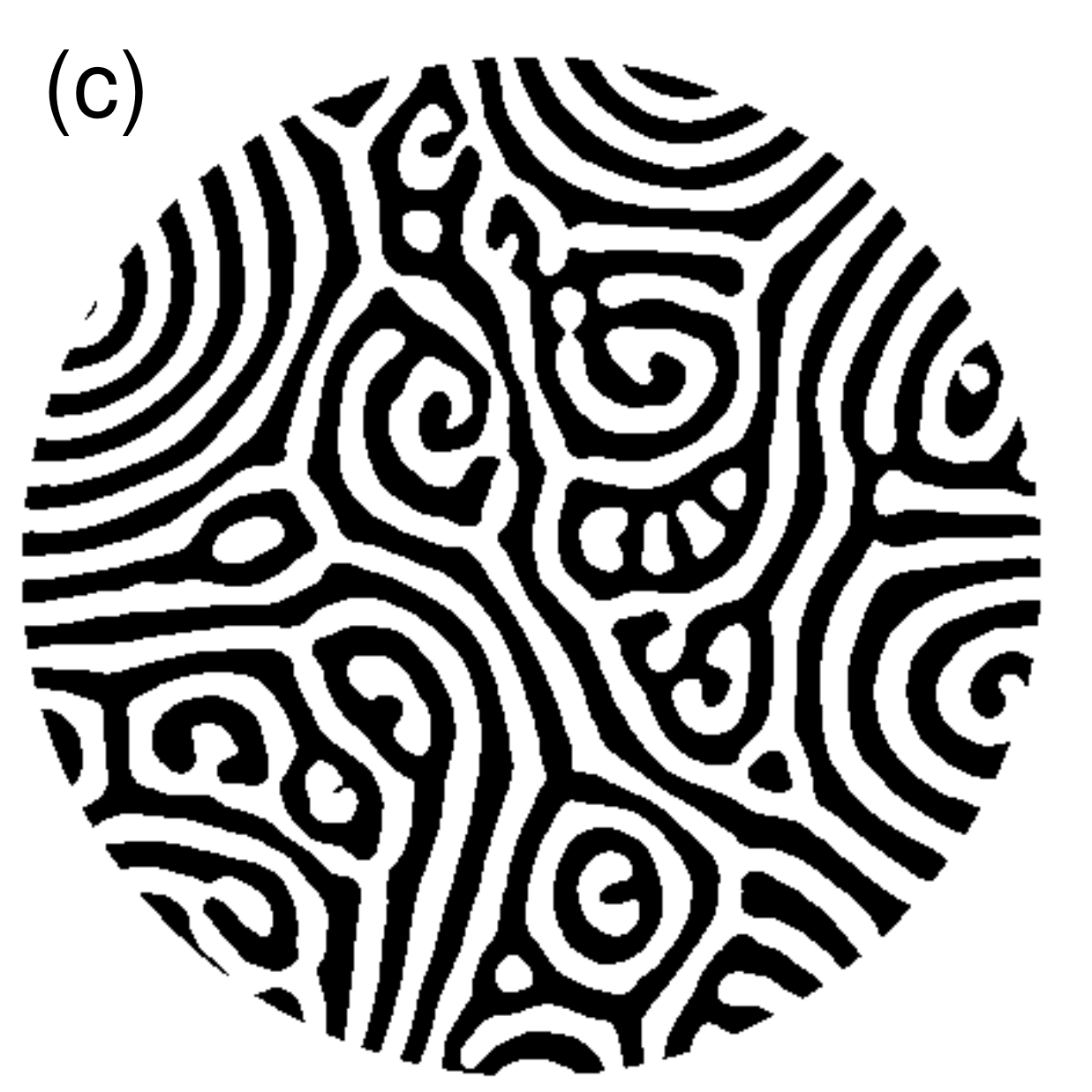}}
{\includegraphics[width=4.2cm]{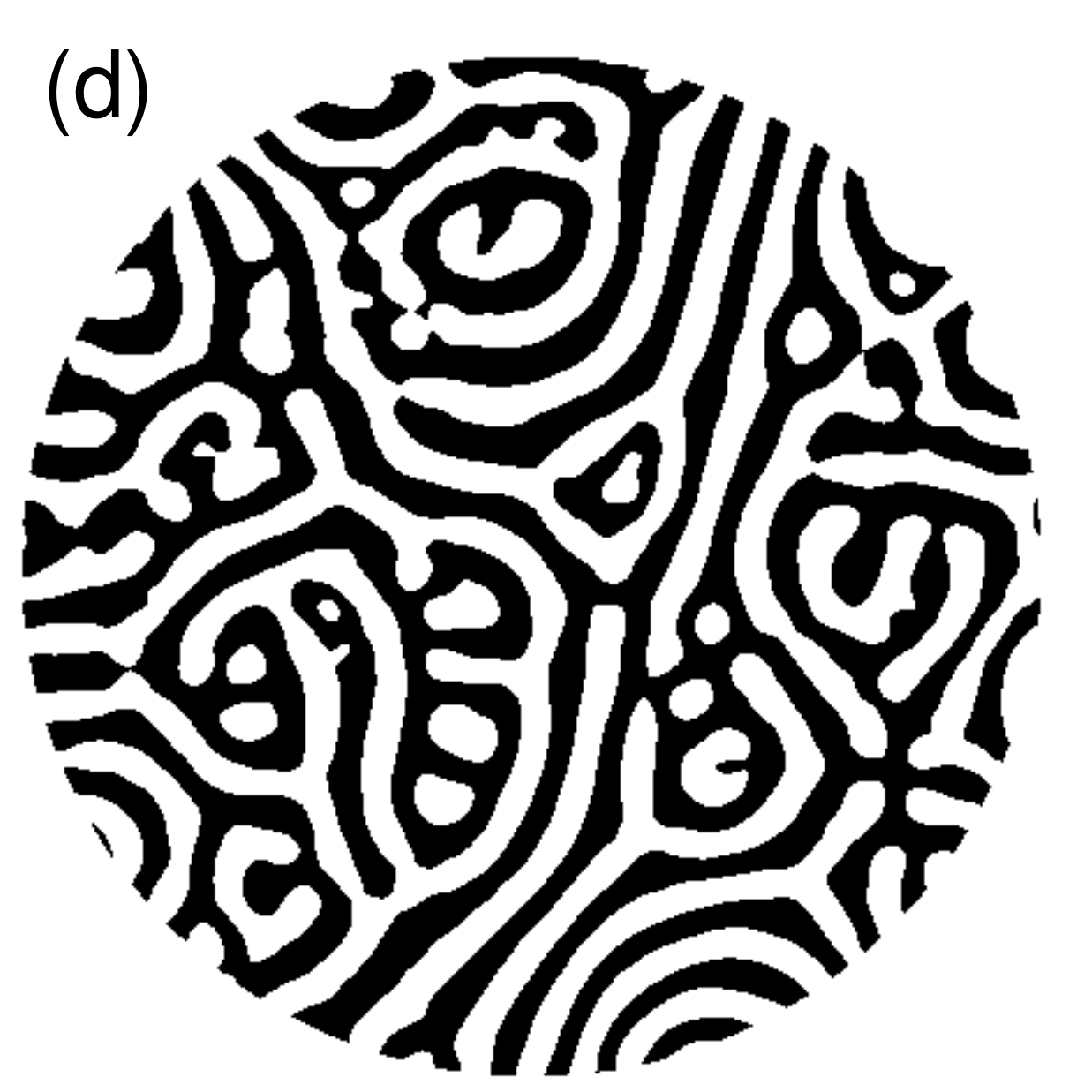}}
\caption{Shadowgraph images of spiral defect chaos convection patterns at reduced
Rayleigh  numbers $\epsilon \approx 1.0$ (a) and $\epsilon \approx 2.0$ (b) are
converted to binary valued images [(c) and (d), respectively] by thresholding the
data at the median value of intensity.  In (c) and (d), the black regions correspond
to hot upflow, and the white regions to cold downflow.}%
\label{sdc_images}%
\end{center}%
\end{figure}

We describe an approach that uses algebraic topology to extract consistent and robust
geometric properties from experimental observations of spatiotemporal complexity. We
employ homology theory since it is the most computable algebraic topological invariant
(i.e., remains constant under deformations and is independent of any particular metric)
that still provides detailed geometric information. Hydrodynamic systems readily produce
complex spatiotemporal patterns \cite{Manneville}, which are ideal for testing these
techniques. We use homology to obtain a limited set of numbers (called {\em Betti numbers}
defined below) that characterize dynamically relevant features in a weakly turbulent state
of Rayleigh-B\'{e}nard convection (RBC) known as spiral defect chaos (SDC) \cite{Morris}
(see Fig.~\ref{sdc_images}). In particular, we find that: (a) the geometric structures of
asymptotic states of SDC are readily distinguishable as a function of a control parameter;
(b) the entropy of the SDC attractors quantifies the evolution of the system through
different geometric configurations; (c) the asymptotic global geometric configurations
evolve stochastically as opposed to chaotically as identified previously \cite{Egolf1,Egolf2};
(d) symmetry breaking in the flow patterns is readily detectable in cases where simple
statistical measures (e.g., measurements of mean pattern volumes) are insensitive to
pattern asymmetries; and (e) the boundary and bulk patterns can be distinguished.

We measure convective flow in a 0.069~cm deep horizontal layer of CO$_2$ gas at a
pressure of 3.2~MPa bounded above by a 5~cm thick sapphire window and below by 1~cm
thick gold plated aluminum mirror. The lateral walls are circular, formed out of an
annular stack of filter paper sheets 3.8~cm in diameter. An electrical resistive heater
is used to heat the bottom mirror, and the top window is cooled by circulating chilled
water at 21.2~$^o$C. When the vertical temperature difference, $T$, across the gas exceeds
a critical temperature difference, $T_c \approx 4$~$^o$C, the onset of fluid motion
occurs.  The flow organizes into a pattern of convective rolls where hot regions of the
gas move upward and cold regions flow downward. The convection rolls become spatially
disordered and exhibit complex time dependence when the system control parameter
$\epsilon = (T-T_c)/T_c$ (the reduced Rayleigh number) is sufficiently large.  For the
present study, the patterns are observed for the order of
$10^4\tau_\nu$ ($\tau_\nu \approx 2.1$ seconds is the vertical thermal diffusion time)
at $\epsilon \approx 1.0$ and $\epsilon \approx 2.0$.

Shadowgraph visualization of the convecting flow yields intensity images, which are
used to compute the topology of the rolls in SDC (Fig.~\ref{sdc_images}).
The shadowgraph technique is sensitive to the refractive index variations in the gas,
which represents the temperature profile of the flow \cite{ShadowBook}.
The state of the flow is sampled at 11 Hz by capturing raw intensity images
[Fig.~\ref{sdc_images} (a) \& (b)] using a 12-bit digital camera; a background image
of the fluid below the onset of convection is then subtracted to remove optical
nonuniformities (e.g., optical imperfections in the bottom plate). Digital Fourier
filtering is then applied to remove high wavenumber components due to camera spatial
noise. The data is then converted to binary values by thresholding at the median value
of intensity. The resulting images [Fig.~\ref{sdc_images} (c) \& (d)], where black
represents hot upflow and  white represents cold downflow, are used as input to
computation of homology. In what follows, $X(n,\epsilon)$ represents the $n^{th}$
binary image in a time series at the reduced Rayleigh number $\epsilon$. Moreover,
$X^{hot}(n,\epsilon)$ and $X^{cold}(n,\epsilon)$ denote the hot flow and cold flow
regions, respectively, of   $X(n,\epsilon)$. It is worth noting that the number of
pixels in $X^{hot}(n,\epsilon)$ is equal to the number of pixels in $X^{cold}(n,\epsilon)$,
i.e., hot and cold regions occupy the same area in the pattern.

\begin{figure}[!hbt]%
\begin{center}%
{\includegraphics[width=4.2cm]{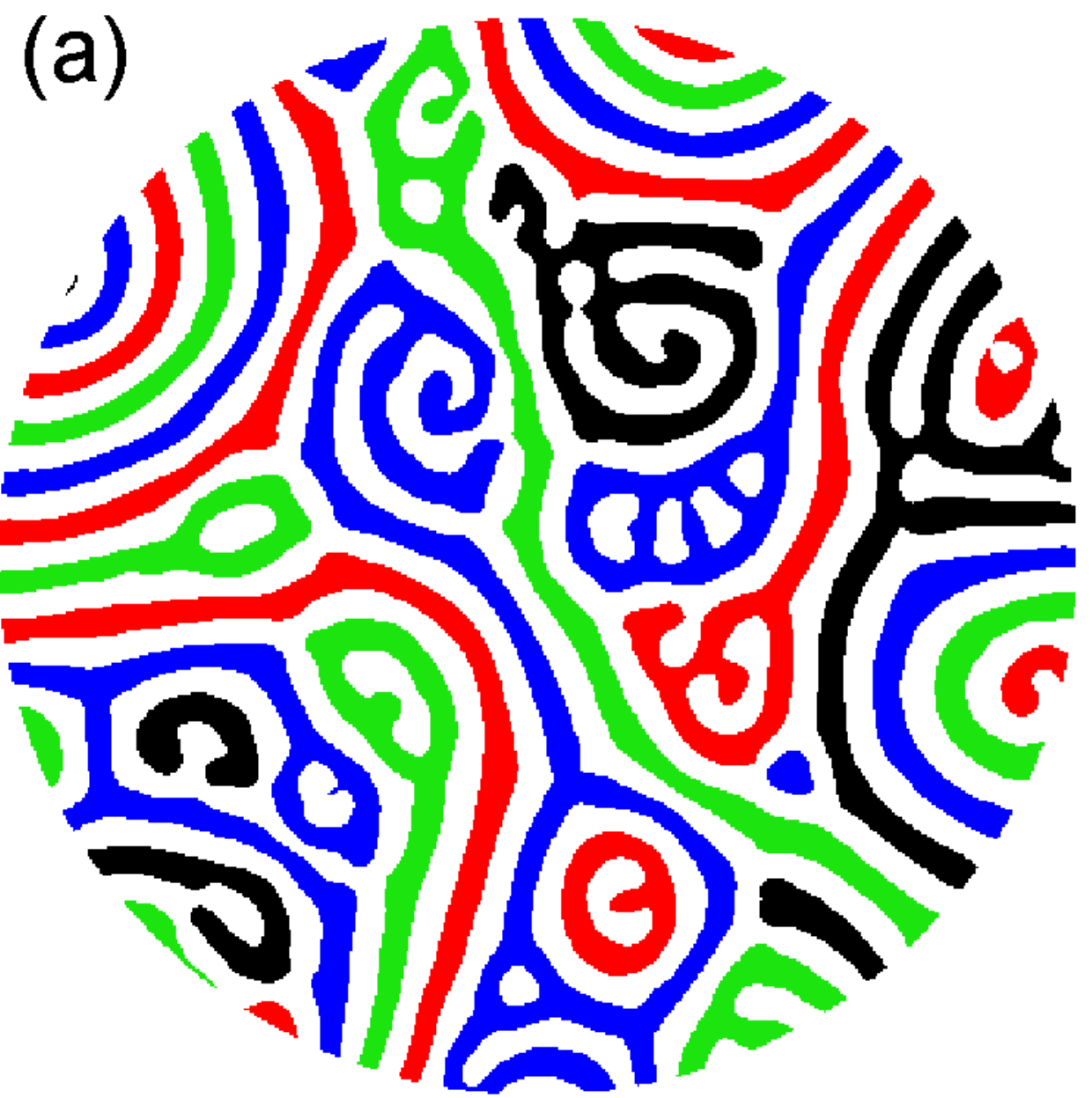}}
{\includegraphics[width=4.2cm]{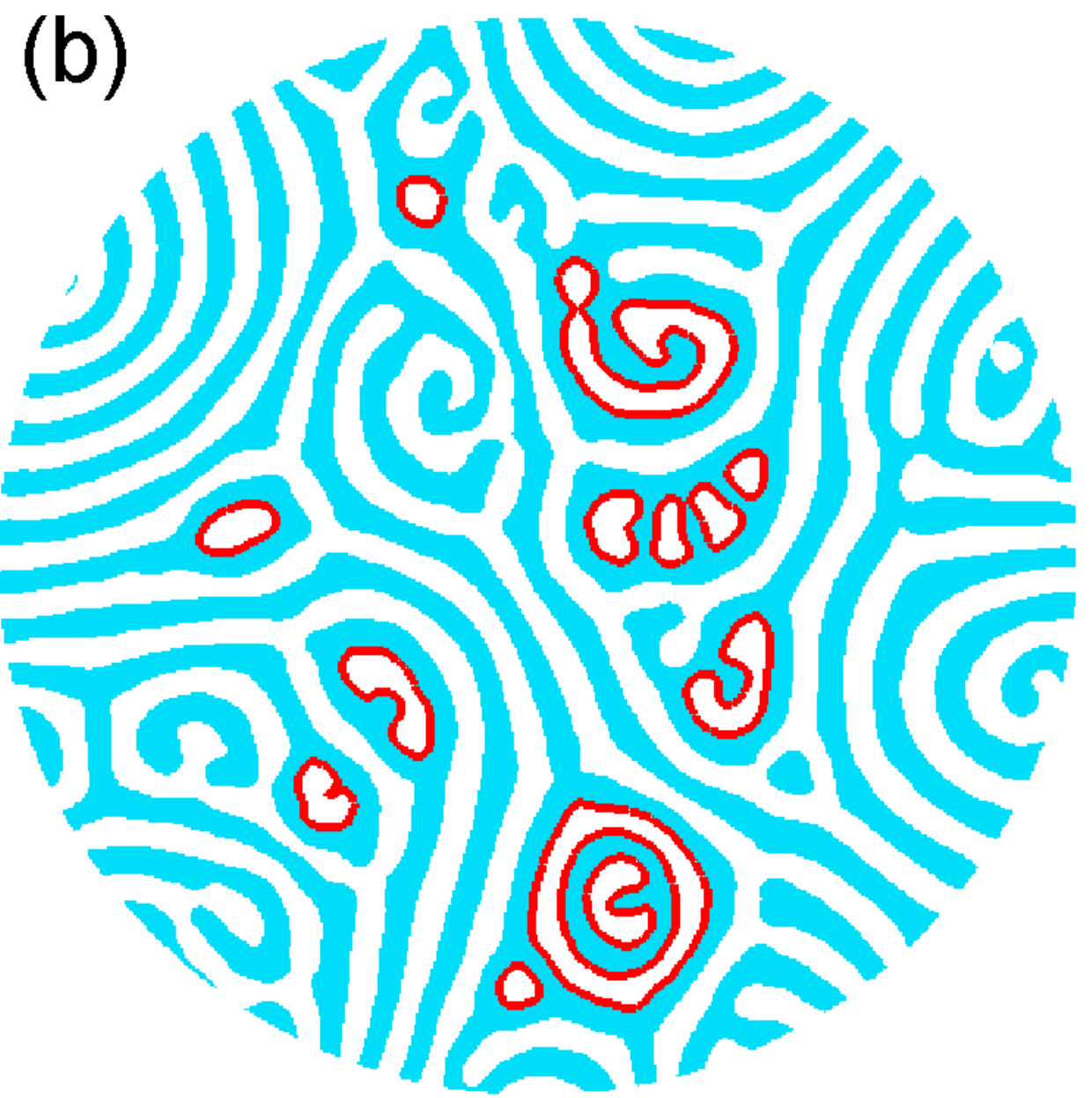}}
\caption{(Color online) Betti numbers for the $X^{hot}$ region shown in
Fig.~\ref{sdc_images} (c). Image (a) shows the $\beta_0^{hot} = 34$ distinct
components, with neighboring components distinguished by different colors.
Image (b) indicates the boundaries of the $\beta_1^{hot} = 13$ holes. The only
components, out of the $\beta_0^{cold} = 38$, of $X^{cold}$ (white region) that
do not touch the boundary are the ones enclosed by the $\beta_1^{hot} = 13$
loops in $X^{hot}$. So the number of components of $X^{cold}$ touching the
boundary is $\beta_0^{cold} -\beta_1^{hot} = 25$.}%
\label{hom_pattern}%
\end{center}%
\end{figure}

Homology theory provides a rigorous, systematic, and dimension independent method for
characterizing geometric structures in $X(n,\epsilon)$ using a few numbers. More
specifically, the homology of a structure $X$ in two dimensions (e.g., $X^{hot}(n,\epsilon)$
or $X^{cold}(n,\epsilon)$) is characterized by two nonnegative integers $\beta_i$, $i=0,1$
called Betti numbers, where $\beta_0$ counts the number of connected components (pieces) of
$X$, and $\beta_1$ is equal the number of holes in $X$. A concrete illustration of this
classification applied to convection patterns is given in Fig.~\ref{hom_pattern}.
(For a more detailed discussion see \cite{Gameiro1,HomBook}.) The computations were done
using the package {\tt CHomP} \cite{Hom,HomBook}. Typically it takes about 10 seconds to
reduce each shadowgraph image (of size 515 x 650 pixels) to binary form and to compute
its homology.

\begin{figure}[!hbt]%
\begin{center}%
{\includegraphics[width=8.5cm]{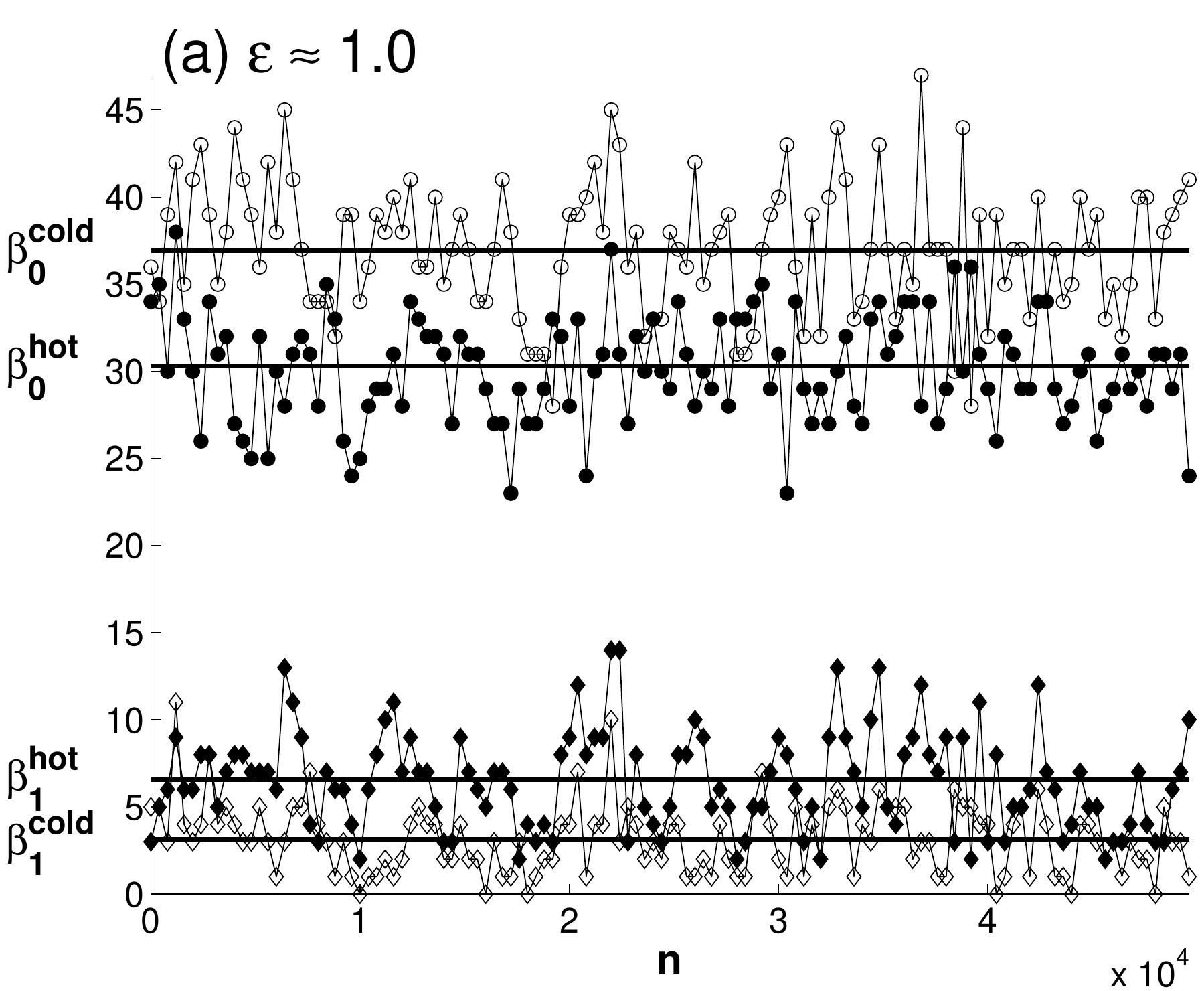}} \\
\vspace{6pt}%
{\includegraphics[width=8.5cm]{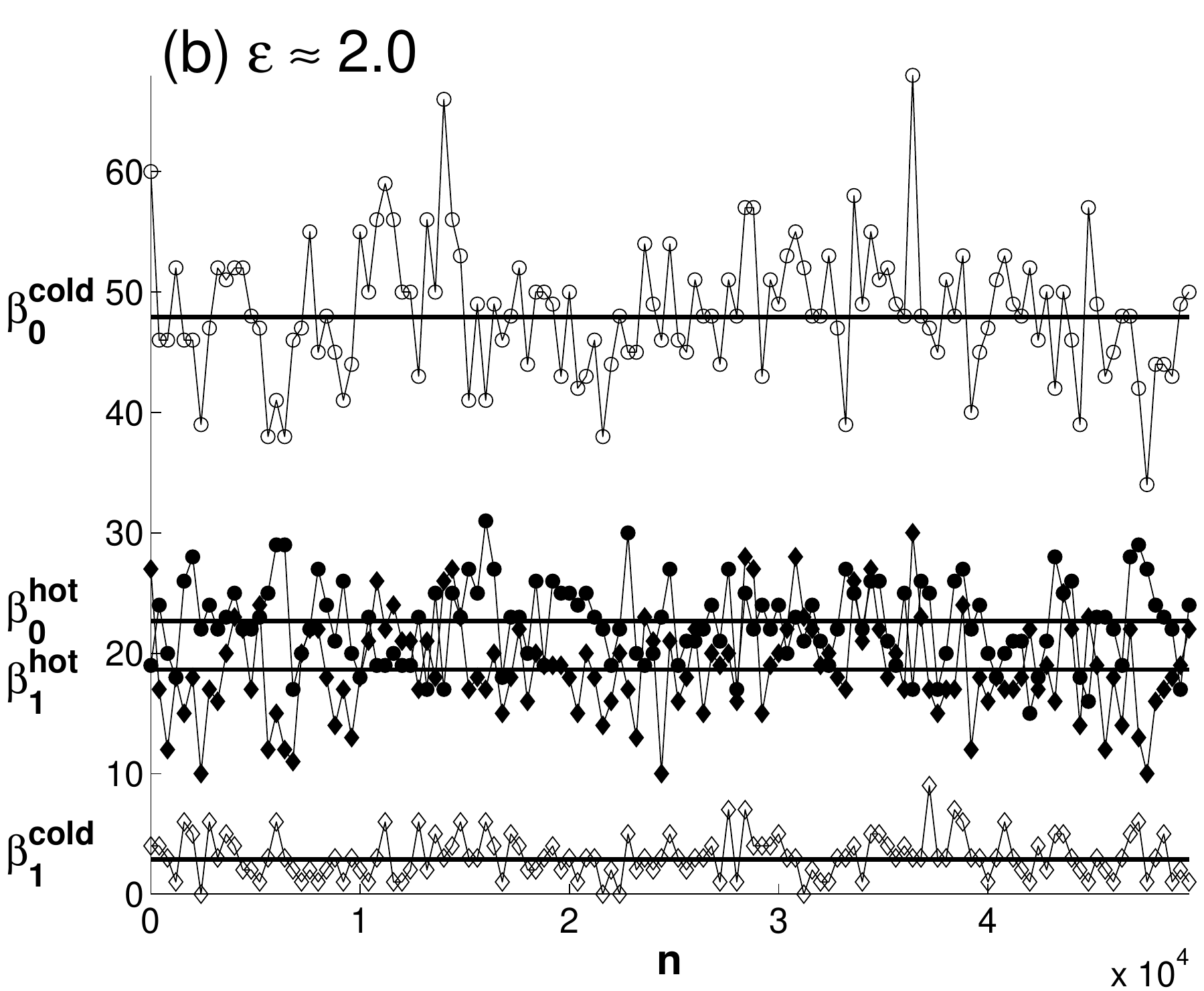}}
\caption{Time series of Betti numbers showing every 400th data point of the
time evolution of the number of distinct components and holes in the regions
$X^{hot}$ (closed, black symbols) and $X^{cold}$ (open, white symbols).
The complete time series plots are qualitatively similar to the ones shown,
the data is partially sampled only for better visualization.
Circles correspond to $\beta_0$ and diamonds to $\beta_1$. The horizontal lines indicate
the mean values $\bar{\beta}_i(\epsilon)$ for the corresponding time series.
Image (a) is for $\epsilon \approx 1.0$ and (b) for $\epsilon \approx 2.0$.}
\label{fig_BettiTs}%
\end{center}%
\end{figure}

{\em Parameter Distinction.---}
The sequence of Betti numbers can be used to clearly distinguish different complex
states of SDC (Fig.~\ref{fig_BettiTs}). For the Rayleigh numbers $\epsilon \approx 1.0$
and $\epsilon \approx 2.0$ we calculated sequences of Betti numbers
$\beta^{hot}_i (n, \epsilon)$ and $\beta^{cold}_i (n, \epsilon)$, $n =0, \ldots, 50,000$.
Figure ~\ref{fig_BettiTs} shows that the time average of the Betti numbers are different
at different $\epsilon$.

The difference in the mean values of the Betti numbers (Fig.~\ref{fig_BettiTs}) reflects
the instability mechanisms operating during the evolution of the complex spatiotemporal
pattern in SDC. The dynamically significant events for the evolution occur in regions of
local compression or dilatation of the roll structure. The compression leads to the merging
of neighboring rolls while the dilatation results in the formation of a new roll in the
pattern. Such local events change the topology of the pattern. For instance, the merging
of two rolls reduces the number of distinct rolls reducing $\beta_0$ by one. A pattern
increasingly dominated by such instabilities would hence show an overall reduction in
$\beta_0$ as different components link together locally at distinct locations. As the
number of distinct rolls reduces, such linkages are often self-intersections of a roll.
This manifests as an increase in the number of loops ($\beta_1$) in the pattern. The
mechanisms characterizing the secondary instabilities for ideal straight rolls \cite{Busse}
have a similar structure and are known as the skew-varicose (compression) and the cross-roll
(dilatation) instabilities. In spiral defect chaos such events are seen to be operational
in regions localized by the curvature of the rolls.

{\em Entropy.---}
At different parameter values, the state of SDC converges to different attractors. The
signature of convergence to the attractor is the evolution of the system in a bounded
neighborhood in the space spanned by the Betti numbers. The set of Betti numbers describes
the different patterns of rolls attained by the system. A complete description of the
homological configuration of the state is determined by the set of Betti numbers describing
the hot and the cold regions.

We calculate the probability of being in a particular configuration from a long time
series of Betti numbers for the evolution of the state while on the attractor. The
probability distribution enables the computation of an entropy to distinguish between
different SDC attractors. The entropy is defined as
$$S(\epsilon)=-\sum_i p_i log(p_i),$$
where the index $i$ span the different states attained by the system as characterized
by the Betti numbers. We denote by $p_i$ the probability of being in a particular state
described by $\beta^{hot}_0$, $\beta^{hot}_1$,$\beta^{cold}_0$ and $\beta^{cold}_1$. The
probability is obtained by trivially counting the number of distinct configurations
attained over a large (ergodic) period of time and normalizing by the total number of
possible configurations. While the entropy is computed over the states attained by the
system in a finite amount of time, its value is seen to asymptote to a constant for long
time series as the probability distribution of the Betti numbers at the attractor converges.
In states where the components do not interact, the entropy reduces as a result of no
fluctuations in the Betti numbers associated with defect formation. The entropy increases
for attractors that show evolving topologies mediated through defects. We find that the
entropy of the system increases from 8.927 to 9.631 as $\epsilon$ is increased from
$\epsilon \approx 1.0$ to $\epsilon \approx 2.0$.

\begin{figure}[!hbt]%
\begin{center}%
{\includegraphics[width=8.5cm]{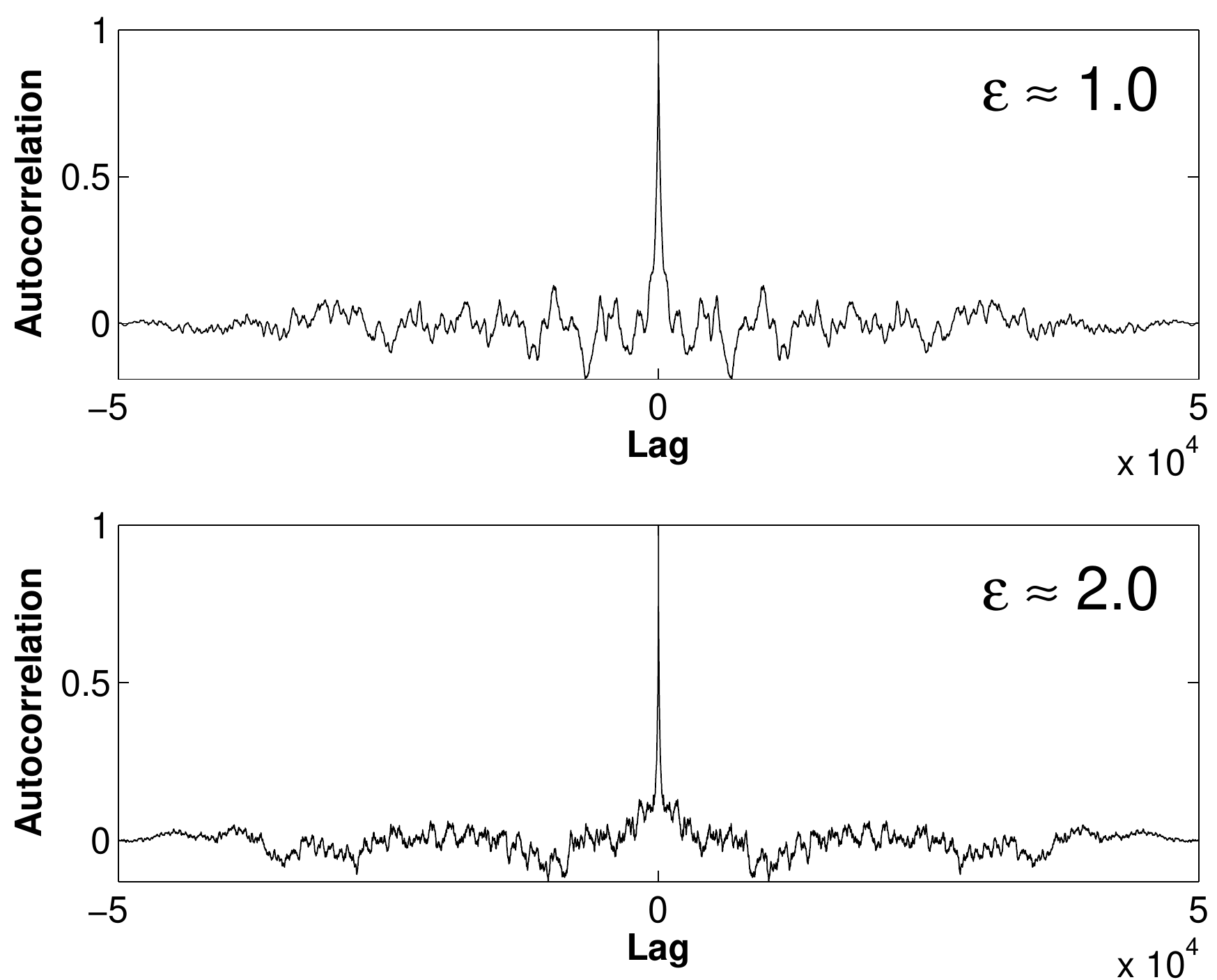}}
\caption{The autocorrelation for the time series of the number of holes ($\beta_1$)
in a state of SDC. The spiked nature of the function suggests that the fluctuations
are primarily stochastic with little correlation. The autocorrelation for $\beta_0$
has a similar spiked shape.}
\label{autocor}%
\end{center}%
\end{figure}

{\em Stochastic Evolution.---}
The evolution of complex geometries may be distinguished as being chaotic or stochastic
from the time series of Betti numbers. The sequence of Betti numbers has been used to
uncover global chaotic evolution through the computation of the largest Lyapunov exponent
in numerical simulations of reaction-diffusion systems previously \cite{Gameiro1}. In our
experiments on SDC we have been unsuccessful in extracting Lyapunov exponents using similar
techniques. In the case of SDC a mechanism describing locally chaotic islands driving the
complex dynamics has been proposed \cite{Egolf1,Egolf2}. It is likely that the
spatiotemporally localized nature of such instabilities in SDC decorrelates at a scale
that is not effectively captured by the time series of Betti numbers; a succession of
these local events interspersed across the system may cause effectively stochastic
evolution for the global geometric structure attained at the attractor. To first order
we find it likely that the dynamics of fluctuations in the Betti numbers may be primarily
stochastic in nature, as it is also suggested by the auto-correlation of the time series
(Fig.~\ref{autocor}).

{\em Symmetry Breaking.---}
The homology of the states exhibited by SDC uncovers a break in the symmetry between
upflows and downflows. For an ideal Rayleigh-B\'{e}nard problem, as described by the
Boussinesq equations \cite{Chandrasekhar}, the inversion of the sign of the velocity
(and time) is also a solution to the evolution equations. This symmetry of the flow
(the Boussinesq symmetry) in an ideal Rayleigh-B\'{e}nard problem suggests that the
statistical properties of the patterns would be the same for $X^{hot}$ and $X^{cold}$;
in particular we should obtain
$\bar{\beta}_i^{hot}(\epsilon) = \bar{\beta}_i^{cold}(\epsilon)$ for $i=0,1$.
As Fig.~\ref{fig_BettiTs} indicates, in our experiments we found that that the mean
values of the Betti numbers clearly distinguish between fluid regions comprised of hot
and cold flows. Furthermore, this asymmetry is enhanced with an increase in the Rayleigh
number. We suspect that the pattern homology serves as a sensitive detector of
non-Boussinesq effects that are present due to the variation in physical properties
of the fluid between the hot bottom layer and the cool top layer. The strength of
non-Boussinesq effects in experiments can be estimated by a dimensionless parameter
$Q$ \cite{Busse,Cross} computed perturbatively at the primary instability for
convection. In our experiments with CO$_2$, we find this parameter to be equal 0.54
near the onset of convection --- thus indicating strong, $O(1)$, non-Boussinesq effects.
A similar computation at $\epsilon \approx 1.0$ and $\epsilon \approx 2.0$ yield $Q$
equaling 1.04 and 1.49 respectively.

\begin{figure}[!hbt]%
\begin{center}%
{\includegraphics[width=8.5cm]{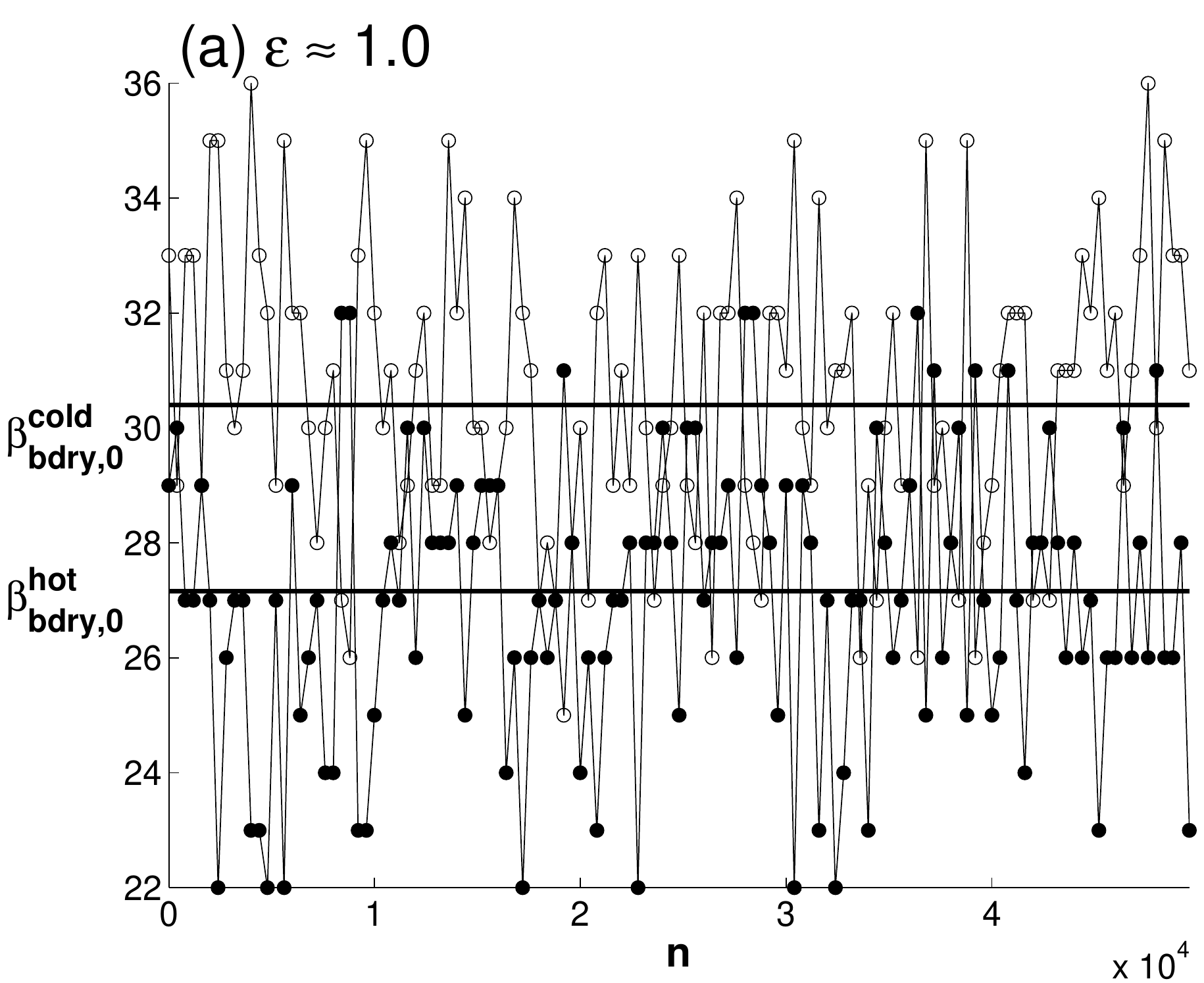}} \\
\vspace{6pt}%
{\includegraphics[width=8.5cm]{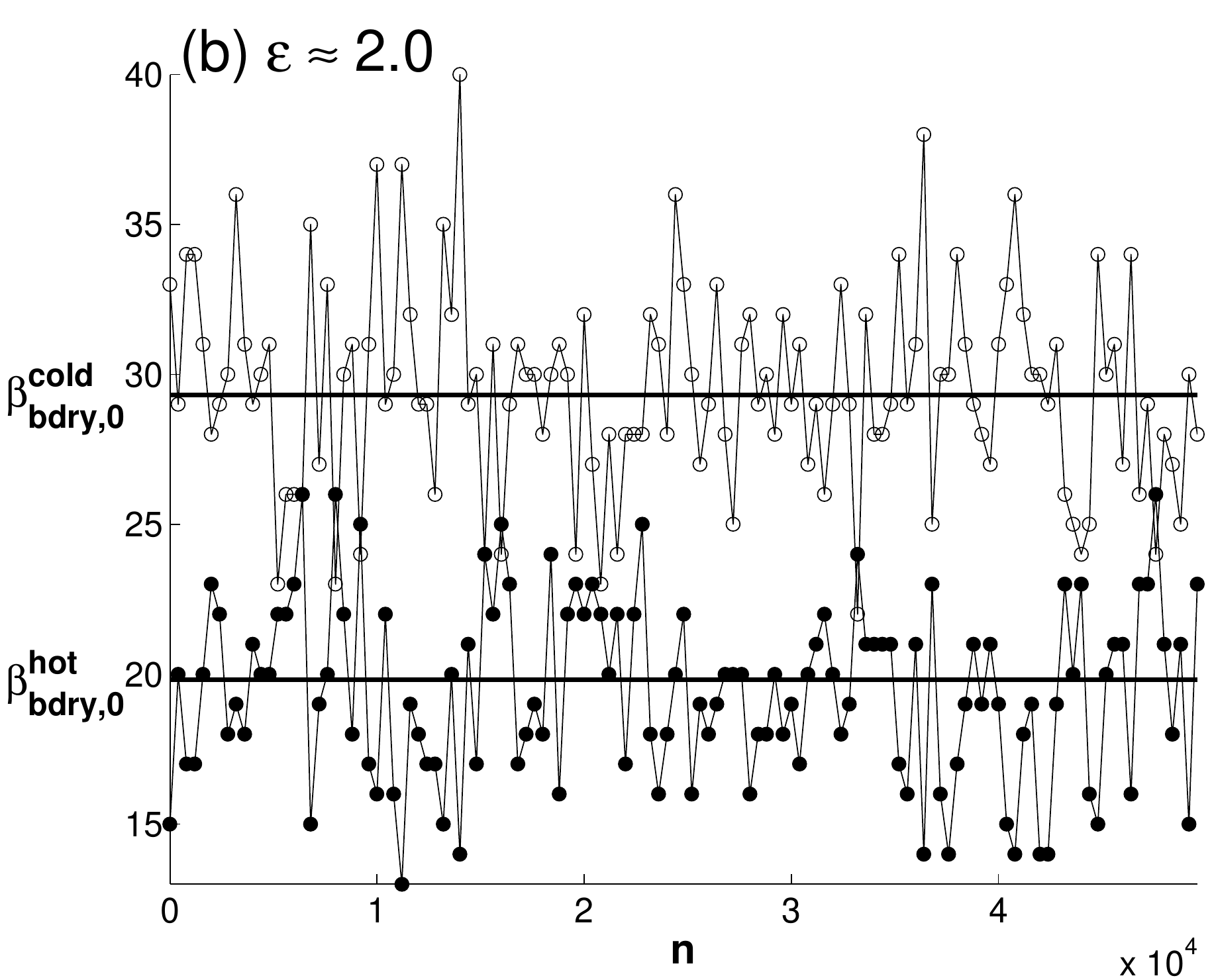}}
\caption{The number of components connected to the boundary for $\epsilon \approx 1.0$
(a) and $\epsilon \approx 2.0$ (b). As in Fig.~\ref{hom_pattern} closed circles
represent $X^{hot}$ and open circles represent $X^{cold}$. A sample from every 400th
data point from the complete time series is displayed in the above figures.}
\label{fig_bdry}%
\end{center}%
\end{figure}

{\em Boundary Effects.---}
The pattern homology can also provide a well-defined way to separate boundary-driven
effects from bulk phenomena in pattern forming systems. Typically, separating boundary
behaviors from bulk effects is done by setting a cutoff based on pattern  correlation
lengths. Using homology, we distinguish regions of a pattern as being part of the bulk
if they are isolated from the boundary. These bulk regions of a given component
(say, for example, isolated hot regions) are easily counted by recognizing they comprise
the interior of holes of the other component (in this example, cold holes, as characterized
by $\beta_1^{cold}$). Thus, the number of components of $X^{hot}$ connected to the boundary
is $\beta_{bdy,0}^{hot} := \beta_0^{hot}-\beta_1^{cold}$ (see Fig.~\ref{hom_pattern}).
Similarly for $X^{cold}$, $\beta_{bdy,0}^{cold} := \beta_0^{cold}-\beta_1^{hot}$. One
notices an asymmetry between the two components in this measure (Fig.~\ref{fig_bdry})
with a smaller number of components connected to the boundary as $\epsilon$ increases for
hot rolls as opposed to being almost the same for cold rolls. This also reflects in the
number of components in the bulk, $\beta_0-\beta_{bdy,0}$, where the $\epsilon$
dependence is primarily seen in the cold rolls.

In summary, we have presented a robust topological characterization of experimental data
from a spatiotemporal complex dynamical system. This technique gives one the ability to
map the gross global configuration of a state to a few dimensionless numbers. We found
stochastic (and ergodic) global topological dynamics for the system of spiral defect chaos
where the underlying microscopic dynamics is deterministic and previously observed to be
locally chaotic. We also found asymmetric topological configurations between hot and cold
regions for which we are unable to provide a physical mechanism. A more complete description
of the system would require coupling such scale-independent measures with other measures
that depend on the length-scales of the patterns, e.g. the mean length of a roll between
defects. Additionally, it would be interesting to compare the correlation of local
characterizations such as Lyapunov exponents or wave numbers with variations in quantitative
topological measures. Understanding the interplay between local and global dynamics is a
core issue in the study of complex systems.

\vspace*{0.2cm}
{\em Acknowledgements.---}
M. G. was partially supported by CAPES, Brazil. K. M. was partially supported by
NSF-DMS-0107395. Support by NSF-ATM-0434193 is gratefully acknowledged by K. K. and M. S.

\bibliography{sdc_prl}

\end{document}